\documentclass[aps,pra,twocolumn,longbibliography,superscriptaddress,letterpaper]{revtex4-1}
\usepackage{graphicx}
\usepackage{amsmath}
\usepackage{amssymb}
\usepackage{esint}
\usepackage{verbatim}
\usepackage{xcolor}
\usepackage{soul}
\usepackage{siunitx}
\usepackage{hyperref}
\usepackage[english]{babel}

\newcommand{\be}{\begin{equation}}
\newcommand{\ee}{\end{equation}}
\newcommand{\ba}{\begin{array}}
\newcommand{\ea}{\end{array}}
\newcommand{\bqa}{\begin{eqnarray}}
\newcommand{\eqa}{\end{eqnarray}}

\begin{document}

\title{Tunneling-induced fractal transmission in valley Hall waveguides}

\author{Tirth Shah}
 \affiliation{Max Planck Institute for the Science of Light, Staudtstrasse 2, 91058 Erlangen, Germany}
\affiliation{Department of Physics, Friedrich-Alexander Universit\"at Erlangen-N\"urnberg, Staudtstrasse 7, 91058 Erlangen, Germany}
 \author{Florian Marquardt}
 \affiliation{Max Planck Institute for the Science of Light, Staudtstrasse 2, 91058 Erlangen, Germany}
 \affiliation{Department of Physics, Friedrich-Alexander Universit\"at Erlangen-N\"urnberg, Staudtstrasse 7, 91058 Erlangen, Germany}
 \author{Vittorio Peano}
 \affiliation{Max Planck Institute for the Science of Light, Staudtstrasse 2, 91058 Erlangen, Germany}
 \email{vittorio.peano@mpl.mpg.de}

\date{\today}

\begin{abstract} 
The valley Hall effect provides a popular route to engineer robust waveguides for bosonic excitations such as photons and phonons. The almost complete absence of backscattering in many experiments  has its theoretical underpinning in a smooth-envelope approximation that neglects large momentum transfer and is accurate only for small bulk band gaps and/or smooth domain walls.  For larger bulk band gaps and hard domain walls backscattering is expected to  become significant. Here, we show that in this experimentally relevant regime,  the reflection of a wave at a sharp corner  becomes highly sensitive on the orientation of the outgoing waveguide relative to the underlying lattice. Enhanced backscattering can be understood as being triggered by resonant tunneling transitions in quasimomentum space. Tracking the resonant tunneling energies as a function of the waveguide orientation reveals a self-repeating fractal pattern that is also imprinted in the density of states and the backscattering rate at a sharp corner.

\end{abstract}

\maketitle

\section{Introduction}

Robust waveguides are an essential component for the transport of classical and matter waves in on-chip devices. Ideally, it would be desirable to realize waveguides in which backscattering is strictly forbidden even after arbitrary (but weak) disorder is introduced. This is, however, possible only in topological systems with broken time-reversal symmetry \cite{hasan_colloquium_2010}. For topological fermionic system, Kramers degeneracy also offers a powerful mechanism to prevent any backscattering in the presence of arbitrary geometrical disorder (as long as the disorder does not break the time-reversal symmetry). On the other hand,  topological bosonic systems \cite{bernevig_quantum_2006,kane_quantum_2005}  without broken time-reversal symmetry are never completely immune to  geometrical disorder, see e.g. \cite{ozawa_topological_2019,shah_topologically_2022}. While it is clear that backscattering is not forbidden in this type of waveguide it is usually difficult to quantify this phenomenon which remains poorly investigated.

A popular approach to engineer robust waveguides without the need of breaking the time-reversal symmetry is based on the valley Hall effect \cite{martin_topological_2008,ju_topological_2015}. This is a symmetry-based approach that has found implementation in a range of experimental platforms ranging from electronic
\cite{martin_topological_2008,ju_topological_2015}, to plasmonic \cite{wu_direct_2017}, photonic \cite{ma_all-si_2016,dong_valley_2017,gao_valley_2017,kang_pseudo-spinvalley_2018,noh_observation_2018,shalaev_robust_2019,zeng_electrically_2020,arora_direct_2021}, and mechanical systems \cite{lu_observation_2017,vila_observation_2017,xia_observation_2018,zhang_topological_2018,zhang_directional_2018,tian_dispersion_2020,ren_topological_2022,ma_experimental_2021,zhang_topological_2021,ma_experimental_2021,xi_observation_2021,zhang_gigahertz_2022}. In this setting, the counter-propagating guided modes are valley polarized, that is, they are localized in non-overlapping regions in the quasi-momentum representations. Thus, backscattering requires large quasi-momentum transfer and is suppressed as long as the wavefunction for the guided modes are smooth on the scale of the underlying lattice potential. This also implies that there is a trade-off between two desirable features: transverse localization of the guided modes and resilience of the transmission against disorder. In is well-known that the resilience to backscattering depends also on the orientation of the waveguide relative to the underlying lattice, see e.g \cite{ma_all-si_2016,wu_direct_2017,lu_observation_2017}. In particular, for implementations based on crystals with a triangular Bravais lattice,  waveguides aligned with a basis lattice vector are less prone to backscattering \cite{ma_all-si_2016}. However, a systematic investigation of the orientation dependence is still missing. 

In a recent work \cite{shah_tunneling_2021}, we have provided an interpretation of the backscattering transitions in  valley Hall waveguides as tunneling transitions in the quasi-momentum space. 
Our previous investigation focused on smooth domain walls. In this work, we move our attention to systems with hard domain walls (i.e. system in which the underlying crystal geometry changes abruptly at the domain wall) and consider a regime where the transverse confinement is relatively strong. In particular, we are interested in  experimentally relevant scenarios in which two or more straight waveguides are connected at sharp corners, cf Fig.1(a). We find that the transport and spectral  properties are highly sensitive on the orientation of the waveguides. Extensive simulations of the density of states (DOS) of straight waveguides as well as of the transmission at a sharp corner  give hints of an underlying fractal dependence of these observables on the energy and the waveguide orientation. We provide an explanation of these empirical observations by analyzing the quasi-momentum paths and energies for resonant tunneling.

The fractal spectrum we analyze is reminiscent of the Butterfly spectrum of  2D charge particles on a lattice  as a function of the magnetic flux originally predicted by Hofstadter \cite{hofstadter_energy_1976} and later observed in a variety of platforms including graphene superlattices \cite{dean_hofstadters_2013,ponomarenko_cloning_2013}, cold atoms in optical lattices \cite{aidelsburger_realization_2013}, and superconducting qubits \cite{roushan_spectroscopic_2017}.  In our work, the orientation of the waveguide relative to the lattice plays a similar role as the magnetic flux. Our work is also loosely  connected to the investigation of fractal spectra in quasi-crystal geometries \cite{zilberberg_topology_2021,kraus_topological_2012,verbin_observation_2013,tran_topological_2015,bandres_topological_2016} . Indeed, the waveguides we investigate can be viewed as 1D quasi-crystals for orientations that are not commensurate to the lattice. However, we emphasize that, here, the fractal structure  we describe is not a property of the spectrum for a single orientation but rather of its variation as a function of the orientation.

\section{Review of the valley Hall physics}
\begin{figure}
\centering
    \includegraphics[width=\columnwidth]{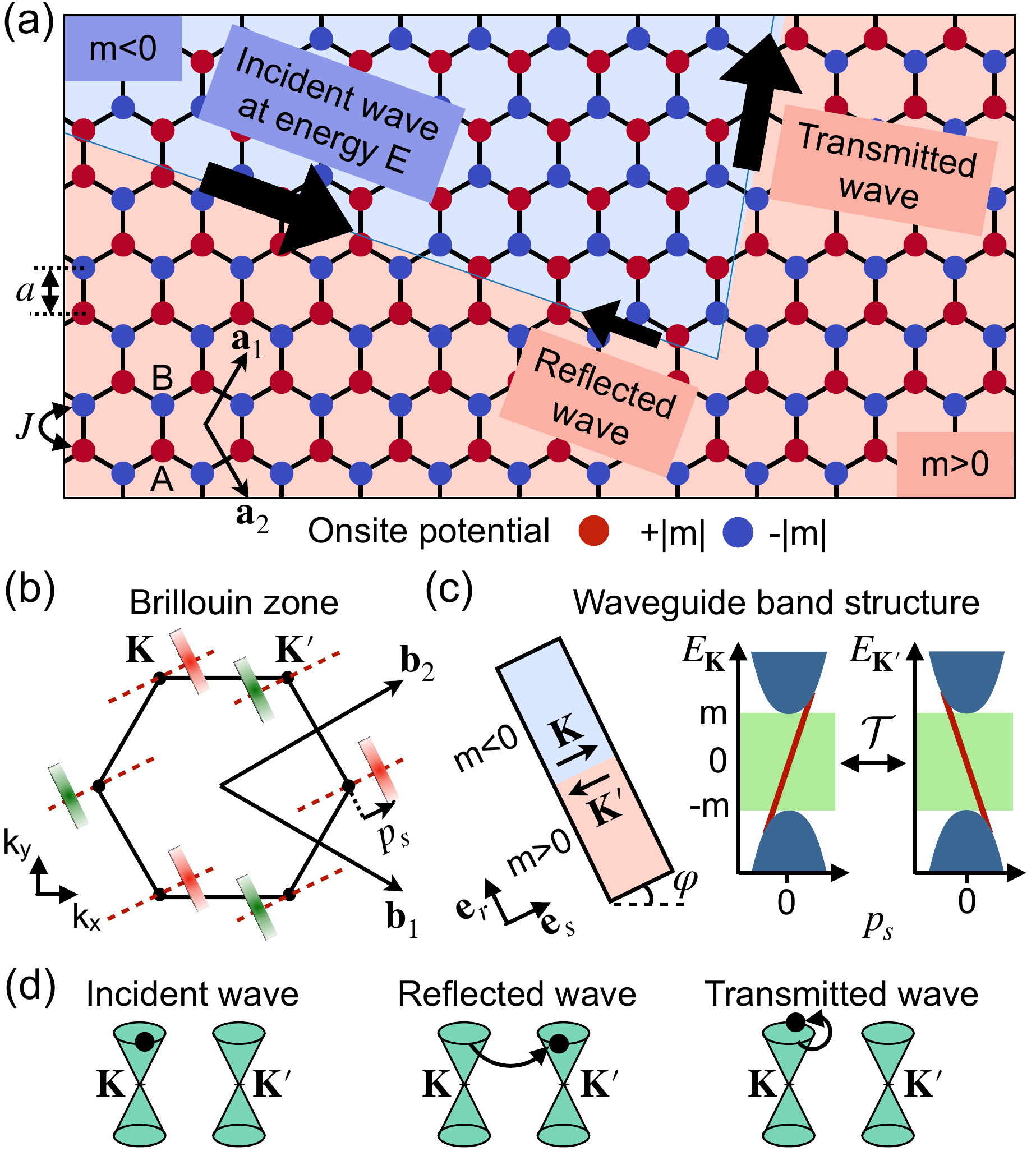}
    \caption{Review of valley Hall effect. 
    (a) Schematic transport of the valley Hall edge channel through a sharp turn. The edge channel travels at the interface of the two distinct domains (red and blue regions), consisting of opposite onsite potentials on the two sublattices A and B. 
    (b) Sketch of the bulk Brillouin zone. Also sketched are the wavefunctions of  two counter-propagating confined modes in the quasi-momentum representation. 
    (c) Mass domain wall defining a straight waveguide. Corresponding valley band structures in the smooth-envelope approximation. 
    (d) Sketch of the Dirac cones. Also shown are the energy and mean quasimomentum  of the confined mode. The reflection of the wave involves a large quasi-momentum transfer to reach a different valley, whereas the transmission is achieved with ease by rotating around the same cone. 
    }
    \label{fig:fig1}
\end{figure}


    
    


In order to set the stage for our work,  we start giving a brief introduction to the valley Hall physics. For this purpose and, more generally, as a case study for our investigation we introduce the simplest toy model that allows to implement valley Hall guided modes. This model  is a simple extension of nearest-neighbors graphene tight-binding model as detailed below. 

Excitations on a honeycomb lattice hops with rate $J$ between nearest-neighbors sites with opposite onsite energies,   $m$ and $-m$ on sublattices A and B, respectively. We allow the amplitude and sign of the  mass parameter $m$ to depend on the unit cell. In this way, the lattice can be viewed as being divided into distinct domains according to the sign of the mass parameter, cf Fig.~\ref{fig:fig1}a. We, thus, arrive to the simple  tight-binding Hamiltonian
\begin{eqnarray}
    &&\hat{H} =  \sum _{\mathbf{x}}m(\mathbf{x}) \left( |\mathbf{x},A\rangle\langle \mathbf{x},A| - |\mathbf{x},B\rangle\langle \mathbf{x},B| \right) \nonumber\\
   && - J\sum _{\langle \mathbf{x},\mathbf{x'} \rangle} |\mathbf{x},A\rangle\langle \mathbf{x'},B| + |\mathbf{x}',B\rangle\langle \mathbf{x},A|.
    \label{eq:real_space_hamiltonian}
\end{eqnarray}
Here,  $\mathbf{x}$ indicates the unit cell position and, thus, it is a lattice vector, $\mathbf{x}=n_1 \mathbf{a}_1 + n_2\mathbf{a}_2$ with $n_{1/2}$  integers and $\mathbf{a}_{1/2}$ unit lattice vectors, $\mathbf{a}_{n}=a\sqrt{3}\left(\sqrt{3}/2, (-1)^{n+1} 1/2 \right)$. In addition, $\langle \mathbf{x},\mathbf{x'} \rangle$ denotes the sum over nearest-neighbors.  
\subsection{Quasi-momentum representation}

To gain insight on the valley Hall physics and in particular the phenomenon of backscattering (discussed below), it is very useful to  introduce the quasi-momentum representation. To switch to the quasi-momentum representation we project the valley Hall Hamiltonian Eq.~(\ref{eq:real_space_hamiltonian}) into a basis of plane waves  $|\mathbf{k}\rangle=A_{\rm BZ}^{-1/2}\sum_{\mathbf{x}} \, e^{i \mathbf{k} \cdot \mathbf{x}}|{\mathbf{x}}\rangle$, 
\begin{equation}
\langle \mathbf{k} |\hat{H}|\mathbf{k} \rangle= m(i\nabla_{\mathbf{k}}) \hat{\sigma}_z + \mathbf{h}(\mathbf{k}) \cdot \hat{\boldsymbol{\sigma}}.
\label{eq:k_space_hamiltonian}
\end{equation}
where $\hat{\sigma}_{\rm i=x,y,z} $ is a set of Pauli matrices acting on the sublattice degree of freedom, $\langle A|\hat{\sigma}_z|A\rangle=1$, $\hat{\boldsymbol{\sigma}}=(\hat{\sigma}_{\rm x}, \hat{\sigma}_{\rm y})$, and $\mathbf{h}=(h_x,h_y)$  is given by
\begin{eqnarray}
    &&h_x=-J-2J\cos\left(\frac{\sqrt{3}k_xa}{2}\right)\cos\left(\frac{3k_ya}{2}\right),\nonumber\\
    &&h_y=-2J\cos\left(\frac{\sqrt{3}k_xa}{2}\right)\sin\left(\frac{3k_ya}{2}\right).  
    \label{fig:eq:h_of_k}
\end{eqnarray}
We note that the quasi-momentum $\mathbf{k}$ is defined modulus a reciprocal lattice vector and can be chosen within the Brillouin zone of area $A_{\rm BZ}=8\pi^23^{-3/2}/a^2$, cf Fig.~\ref{fig:fig1}(b). Thus, the normal modes $\tilde{\boldsymbol{\psi}}(\mathbf{k})$ (in quasi-momentum space) are periodic eigenstates of Eq.~(\ref{fig:eq:h_of_k}),  $$\tilde{\boldsymbol{\psi}}(\mathbf{k}+n_1\boldsymbol{b}_1+n_2\boldsymbol{b}_2)=\tilde{\boldsymbol{\psi}}(\mathbf{k})$$
with $\mathbf{b}_1$ and $\mathbf{b}_2$ unit vectors of the reciprocal lattice.

\subsection{Dirac Hamiltonian and valley Chern numbers}
For the bulk case, $m(\mathbf{x})=m_{\rm bk}$, the quasi-momentum $\mathbf{k}$ is a constant of motion and $\langle \mathbf{k} |\hat{H}|\mathbf{k} \rangle$ reduces to a simple $2 \times 2$ matrix. Its spectrum as a function of the quasi-momentum defines the band structure  given by
\begin{equation}
E_{\rm bulk}(\mathbf{k})=\pm \sqrt{m_{\rm bk}^2+|\mathbf{h}(\mathbf{k})|^2}   \end{equation}
In the special case $m_{\rm bk}=0$, it features  Dirac cones centered at the Dirac high-symmetry  points  $\mathbf{K} = (4\pi/(3\sqrt{3}a),0)$ and $\mathbf{K}' = -(4\pi/(3\sqrt{3}a),0)$. A finite mass $m_{\rm bk}$ opens a band gap of width $2|m_{\rm bk}|$ between the two bands. 
We note that the band structure is independent of the sign of the mass parameter $m$. However, this quantity is still very important in that it determines the symmetry-properties of the underlying normal modes.

Next, we focus our attention on exitations whose  wavefunction $\tilde{\boldsymbol{\psi}}(\mathbf{k})$ is  localized in a so-called valley, i.e. the vicinity of a Dirac point. 
The dynamics of these {\it Valley-polarized} excitations are captured by a Dirac equation obtained by
 expanding the Hamiltonian Eq.~(\ref{fig:eq:h_of_k}) about the relevant  Dirac point. For the $\mathbf{K}$-valley, the Dirac equation reads
\begin{equation}
  \langle \mathbf{K} +  \mathbf{p} |\hat{H}|\mathbf{K} +  \mathbf{p}\rangle \approx m(i\nabla_\mathbf{k}) \hat{\sigma}_z + v  \mathbf{p} \cdot \hat{\boldsymbol{\sigma}}.
    \label{eq:dirac_hamiltonian}
\end{equation}
Here, $\mathbf{p}=\mathbf{k}-\mathbf{K}$ is the quasi-momentum counted off from the high-symmetry point $\mathbf{K}$. We note in passing that the Dirac equation can also be derived from symmetry consideration  in the framework of a smooth-envelope approximation  without any detailed knowledge of the underlying microscopic model, see e.g. \cite{shah_tunneling_2021}.
For the bulk case $m=m_{\rm bk}$, it is possible to associate to the Dirac Hamiltonian a topological invariant the so-called valley Chern numbers $C_V$ \cite{martin_topological_2008}. The valley Chern number assumes two possible half-integer values $1/2$ or $-1/2$, with opposite signs in the two valleys. For the lowest band and the $\mathbf{K}$-valley, we have $C_V={\rm sign}(m_{\rm bk})/2$. Under the assumption that the Berry curvature is peaked in the region of validity of the Dirac equation (which is true as long as $m_{\rm bk}/J\ll 1$), the valley Chern numbers accurately quantifies the contribution from a valley to the overall band Chern number. 

\subsection{Topological guided modes}

When a domain with positive mass is joined to a domain with negative mass,  cf Fig.~\ref{fig:fig1}(a), the valley Chern number across the surface changes by one unit. In this scenario, the bulk boundary correspondence separately applied to each valley  (with the implicit assumption that the two valleys are not coupled),  predicts the appearance of a pair of guided counter-propagating gapless  modes (one mode for each valley). Thus, we can view such a domain wall configuration as a topological waveguide.

The appearance of guided modes at the interface of two regions with opposite sign of the mass parameter $m$ can also be substantiated by directly diagonalizing the Dirac Hamiltonian (\ref{eq:dirac_hamiltonian}). Indeed, the existence of a guided mode  for the Dirac equation in the presence of a straight domain  wall has been originally predicted by Jackiw and Rebbi \cite{jackiw_solitons_1976}. If we choose a set of coordinates $x_r$ and $x_s$ such that $m(\mathbf{x})=m(x_r)$ with the domain wall on the $x_r=0$ axis and $m(x_r)>0$ in the lower-half plane, cf Fig.~\ref{fig:fig1}(c), the guided mode has  energy dispersion
\begin{equation}
E_{\mathbf{K},p_s} =   vp_s. 
    \label{eq:edge_state_energy}    
\end{equation}
Here, $v=3Ja/2$ is the Dirac velocity and $p_s$ is the component of the quasimomentum  longitudinal to the domain wall, $p_s=\mathbf{p}\cdot \mathbf{e}_s$ where $\bf{e}_s$  is the unit vector in the longitudinal direction.  
State of the art experimental realizations normally feature hard domain walls, $m(\mathbf{x})=-|m_{\rm bk}|{\rm sign}(x_r)$ where $x_r$ is the coordinate transverse to the domain wall, cf Fig.~\ref{fig:fig1}(c). In this case, the Jackiw and Rebbi solution  
has wavefunction 
\begin{eqnarray}
  \boldsymbol{\psi}_{\mathbf{K},p_s}(\mathbf{x}) =  e^{ip_s x_s} e^{i\mathbf{K}\cdot \mathbf{x}} e^{-\vert m_{\rm bk} x_r /v \vert}
    \begin{pmatrix}
    e^{-i\varphi/2} \\ e^{i\varphi/2}
    \end{pmatrix}.
    \label{eq:edge_state_wavefunction}
\end{eqnarray}
Here, 
 $\varphi$ is the angular-coordinate of the mass domain wall, cf Fig.~\ref{fig:fig1}(c). 
The guided mode for  the valley $\mathbf{K'}$ is obtained by applying the time-reversal operator (in this case complex conjugation) to Eq.~(\ref{eq:edge_state_wavefunction}), $\boldsymbol{\psi}_{\mathbf{K'},-p_s}(\mathbf{x})=\boldsymbol{\psi}^*_{\mathbf{K},p_s}(\mathbf{x})$. 

It is instructive to rewrite the guided mode wavefunction in the quasimomentum representation $\Tilde{\boldsymbol{\psi}}_{\mathbf{K},p_s}(\mathbf{k}) = A_{\rm BZ}^{-1/2} \int  d\mathbf{x} \, e^{-i \mathbf{k} \cdot \mathbf{x}} \boldsymbol{\psi}_{\mathbf{K},p_s}(\mathbf{x})$. We find 
\begin{eqnarray}
    && \Tilde{\boldsymbol{\psi}}_{\mathbf{K},p_s}(\mathbf{k}) = \delta (k_s - \mathbf{K}\cdot{\bf{e}_s} - p_s) \, \boldsymbol{\phi}_{p_s}(k_r), \nonumber
    \\
    && \boldsymbol{\phi}_{p_s}(k_r) =  \frac{2m_{bk}/v}{(m_{bk}/v)^2 + (k_r - \mathbf{K}\cdot\bf{e}_r)^2} 
    \begin{pmatrix}
    e^{-i\varphi/2} \\ e^{i\varphi/2}
    \end{pmatrix}. 
    \label{eq:fourier_transform}
\end{eqnarray}
Here, $\bf{e}_r$ is the unit vector in the radial direction, cf Fig.\ref{fig:fig1}(c). Thus, the right propagating guided mode is localised about the quasimomentum
$
 \bar{\mathbf{k}}_{\mathbf{K},p_s}=\mathbf{K}+ p_s\boldsymbol{e}_s   
$ with standard deviation $\delta k_r=2m_{bk}/v$  in the transverse direction. On the other hand, the time-reversed solution in the $\mathbf{K}$'-valley will be localized about  $\bar{\mathbf{k}}_{\mathbf{K}',-p_s}=\mathbf{K}'- p_s\boldsymbol{e}_s=-\bar{\mathbf{k}}_{\mathbf{K},p_s}$,   see the geometrical illustration in Fig.~\ref{fig:fig1}(b).  We note that for $m_{bk}/J\ll 1$, $\delta k_r$ is much smaller than the distance between the two valleys. 

We now consider a setup where two straight waveguides have been connected by a sharp corner, cf Fig.~\ref{fig:fig1}(a).
If  the condition $m_{bk}/J\ll 1$ is fulfilled, the quasimomentum transfer after  turning the corner is much smaller than the quasimomentum transfer that would be required for backscattering, cf Fig.~\ref{fig:fig1}(d). This leads to  suppression of  backscattering and, thus, robust transport. 

We note  that 
 there is a trade off between the bandwidth  for the guided modes and how well backscattering is suppressed (or, equivalently,  how  strongly the guided modes are localized in quasimomentum) \cite{shah_tunneling_2021}. In order to boost the bandwidth, many experiments are realized  in a regime of intermediate values of the dimensionless bulk mass parameter $|m_{\rm bk}|/J$ in which the coupling between the two valleys can not be safely neglected, see e.g. Shaley et al.~\cite{shalaev_robust_2019}, Zeng et al.~\cite{zeng_electrically_2020}, Ren et al.~\cite{ren_topological_2022}, Ma et al.~\cite{ma_experimental_2021}, and Arora et al.~\cite{arora_direct_2021} with $|m_{\rm bk}|/J=0.2, 0.6, 0.6, 0.4, 0.2$, respectively. This has motivated us to investigate the regime of intermediate masses directly diagonalizing numerically Eq.~(\ref{eq:real_space_hamiltonian}) and, thus, accounting also for the coupling between the two valleys. In all simulation below, we have used the dimensionless bulk mass parameter $|m_{\rm bk}|/J=0.3$. We expect qualitatively similar results for similar values of $|m_{\rm bk}|/J$.

\begin{figure*}
\centering
\includegraphics[width=2\columnwidth]{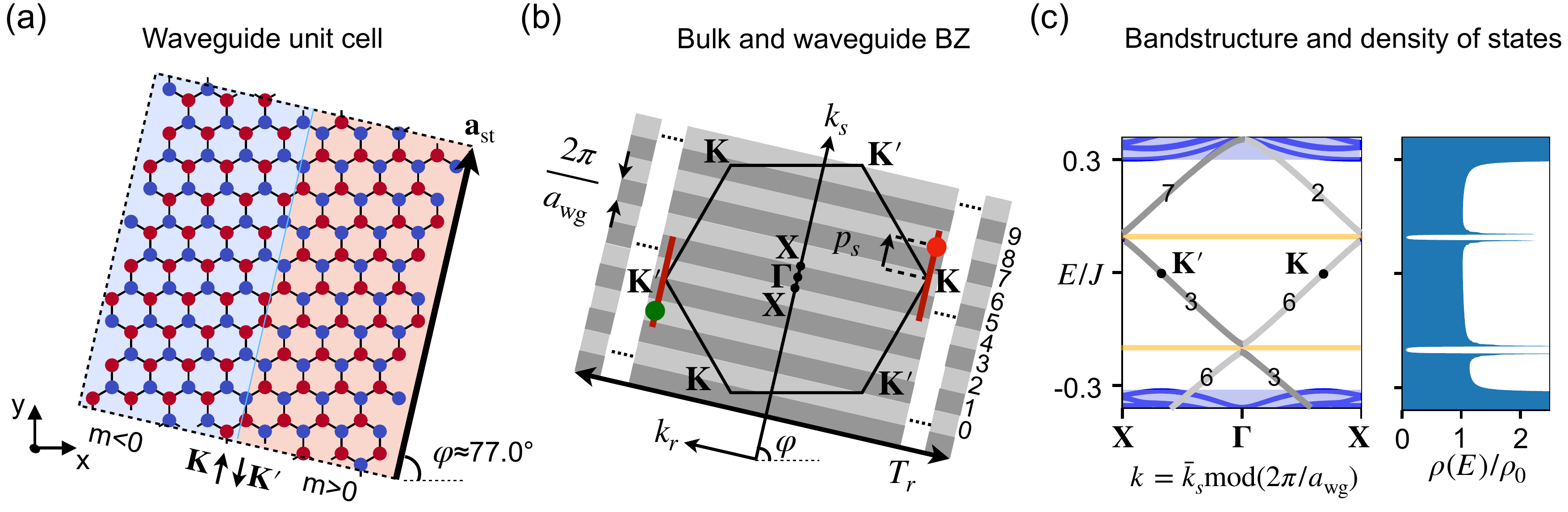}
\caption{Valley Hall effect for a straight waveguide. 
(a) Waveguide unit cell with orientation $\varphi \approx 77.0^\circ $ ($p=2, q=5$). 
(b) Representation of the waveguide BZ (grey stripes) in the reciprocal space. Within the smooth-envelope approximation, the mean quasimomentum of the confined mode  is located along the line (red) passing through the $\mathbf{K}$ or  $\mathbf{K'}$ point and parallel to the domain wall. 
(c) Numerically evaluated band structure and (d) the density of state (normalized to that for the gapless edge channel). The numbers on the edge-band indicate the waveguide BZ index in (b).
}
\label{fig:fig2}
\end{figure*}

\section{Tunneling induced band gaps in valley Hall waveguides}
\label{sec:mingaps}

The simplest signature of the coupling between the two valleys (beyond the Dirac equation approach) is the appearance of minigaps in the  band structure of a straight waveguide. These band gaps have been investigated in \cite{shah_tunneling_2021} with a special focus on the case of smooth mass domain walls $m(\mathbf{x})$. This work has also shown that there is a deep connection between  these band gaps and backscattering in a closed domain wall. For this reason,  we also start investigating the band structure for straight waveguides before switching to the  problem of backscattering at sharp corners  in Section \ref{sec:fractal_backscattering}. Our investigation is complementary to \cite{shah_tunneling_2021} in that we focus on the case of sharp domain walls and corners.

In the Dirac Hamiltonian approach, the momentum $\mathbf{p}$ counted off from the corresponding high-symmetry point is formally treated as a continuous variable, $\mathbf{p}\in \mathbb{R}^2$. This leads to quantized valley Chern numbers and a simple description with a continuous dependence of energy and wavefunction on the angular coordinate $\varphi$, cf Eqs.~(\ref{eq:edge_state_energy},\ref{eq:edge_state_wavefunction}). 
We note, however, that the longitudinal quasi-momentum $p_s$ is not a conserved quantity. A full numerical evaluation (beyond the Dirac equation) of a waveguide's band structure can be straightforwardly performed whenever the waveguide is translationally invariant. This allows to define 
a waveguide lattice constant  $ a_{\rm wg}$,  cf Fig.~\ref{fig:fig2}(a),  and  the corresponding conserved waveguide quasimomentum $k=(\mathbf{k}\cdot\mathbf{e}_s){\rm mod}(2\pi/ a_{\rm wg})$, Fig.~\ref{fig:fig2}(b).  In particular,  the Jackiew and Rebbi solution in the valley $\mathbf{K}$ has waveguide quasimomentum
$k_{\mathbf{K},p_s}=(\mathbf{K}\cdot\mathbf{e}_s+p_s){\rm mod}(2\pi/a_{\rm wg})$ while its time-reversed partner in the $\mathbf{K}'$ valley has opposite  quasimomentum, $k_{\mathbf{K}',p_s}=(\mathbf{K}'\cdot\mathbf{e}_s-p_s){\rm mod}(2\pi/a_{\rm wg})=-k_{\mathbf{K},p_s}$. Thus, the two solutions, which have the same energy, have also the same  waveguide quasi-momentum  at the time-reversal invariant  high-symmetry points $k=\Gamma=0$ or $k=X=\pi/a_{\rm wg}$. At these points, they are resonantly coupled leading to a gap in the guided mode band structure. The band gap can be viewed as a tunnel splitting proportional to the overlap of the two guided modes $\tilde{\boldsymbol{\psi}}_{\mathbf{K},p_s}(\mathbf{k})$, and $\tilde{\boldsymbol{\psi}}_{\mathbf{K}',-p_s}(\mathbf{k})=\tilde{\boldsymbol{\psi}}^*_{\mathbf{K},p_s}(-\mathbf{k})$ which are peaked in different valleys \cite{shah_tunneling_2021}. In this framework, the peaks quasimomenta   $\bar{\mathbf{k}}_{\mathbf{K},p_s}$ and $\bar{\mathbf{k}}_{\mathbf{K}',-p_s}$  (see definition above) can be viewed as the classical quasimomenta of the two counter-propagating guided modes (a rigorous WKB approach is possible for smooth domain walls \cite{shah_tunneling_2021}). The resulting band structure, for the special case $\varphi \approx 77.0^\circ $, is shown in Fig.~\ref{fig:fig2}(c). An alternative representation of the spectrum is provided by the density of states (DOS) $\rho(E)$, Fig.~\ref{fig:fig2}(d). For an infinite size system,  the DOS in the region of the bulk band gap is flat away from the guided modes band gaps, vanishes inside the band gaps, and displays Van Howe singularities at the band edges, cf Fig.~\ref{fig:fig2}(d). Numerically, we suppress finite size artifacts by simulating a large system size in the presence of a small homogeneous broadening of the energy levels, see Appendix \ref{app:density_of_states}.

Until now we have focused  on translationally invariant waveguides. Thus, we have implicitly assumed that the mass domain wall  is aligned with at least one lattice vector $\mathbf{x}$. This is the case  only  for angles $\varphi$ fulfilling the rationality condition \cite{charlier_electronic_2007}
\begin{equation}
    \alpha\equiv\sqrt{3} \cot \varphi = p/q,
    \label{eq:alpha}
\end{equation}
with $p$ and $q$ relative prime integers. The waveguide lattice constant is then  
\begin{equation}
a_{\rm wg}=\frac{3qa}{1+(pq \rm mod 2)\sin \varphi}.    
\end{equation}
We note that the lattice constant  is a discontinuous function of $\varphi$ via $p$ and $q$. Thus, the number of bands and band gaps will also be a discontinuous function of $\varphi$ leading to an overall discontinuous spectrum.  This is reminiscent  of the  discontinuous dependence of the energy spectrum of electrons in an external magnetic field as a function of the  magnetic flux \cite{hofstadter_energy_1976}. In the latter scenario, the appearance of an ever larger number of band gaps for large  denominators $q$ of the magnetic flux leads to  the celebrated Hofstadter butterfly spectrum \cite{hofstadter_energy_1976}. This spectrum has been intriguing for generation of researchers and even the general public  because of its self-similar fractal nature.  This has motivated us to explore numerically the edge state spectrum as a function of the angular-coordinate $\varphi$.


\begin{figure}
\centering    
    \includegraphics[width=\columnwidth]{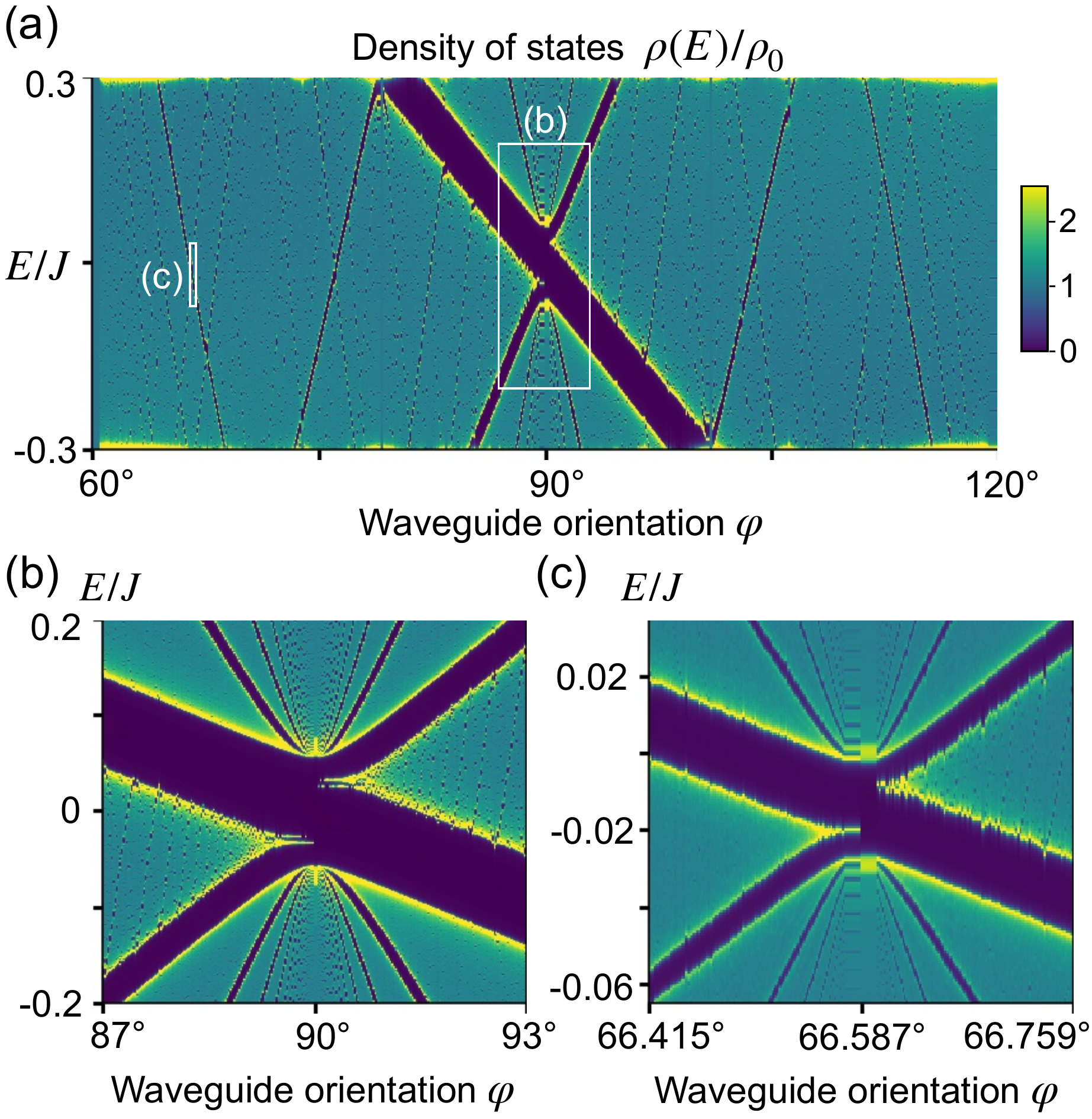}
    \caption{Orientation dependence of the waveguide DOS. 
    (a) Numerically evaluated density of states of the waveguide as a function of energy (in the region of the bulk band gap) and the waveguide orientation  (in the interval $60^\circ\leq \varphi <120^\circ$). 
    (b,c) Zoom-ins of the regions inside the white boxes in (a). The same pattern appears at two different scales. 
    }
    \label{fig:fig3}
\end{figure}

\section{Fractal orientation-dependence of the band structure}
\label{sec:fractal_edge_band_gaps}


The DOS of the waveguide as a function of energy and  orientation $\varphi$  is shown in  Fig.~\ref{fig:fig3}(a). We focus on the interval $60^\circ\leq\varphi< 120^\circ$ as the DOS for angles outside of this interval  can be recovered by applying the following symmetry considerations: (i)  A rotation by $120^\circ$ of the domain wall is equivalent to a rotation of the whole Hamiltonian and, thus, does not change the spectrum, $\{E\}_\varphi=\{E\}_{\varphi+120^\circ}$ (ii) A rotation by $180^\circ$ of the domain wall just changes the sign of all onsite energies. Since the lattice is bipartite each energy level has a corresponding level with opposite energy in the rotated structure, $\{E\}_\varphi=\{-E\}_{\varphi+180^\circ}=\{-E\}_{\varphi+60^\circ}$. As it should be expected given the discontinuous $\varphi$-dependence of the number of band gaps, the density of states displays an intricate structure. A multitude of band gaps (dark blue strips) of varying width  merge at the intersection of the $E=0$ and $\varphi=90^\circ$-axes with the two main band gaps forming an x-shaped pattern, cf also the zoom in Fig.~\ref{fig:fig3}(b).  Surprisingly further magnification about the intersection of the $E=0$ axis and  $\varphi\approx 66.5^\circ$  reveals that the same remarkable pattern is repeated for smaller energy and angle variations, cf  Fig.~\ref{fig:fig3}(c). This raises the question whether further magnification would reveal a self-repeating fractal pattern, analogous to the celebrated Hofstadter's Butterfly spectrum  but now with the angle $\varphi$ playing the role of the magnetic flux. We investigate this question in the next section.  

We note in passing that for any fixed irrational $\alpha$ the waveguide can be viewed as a quasicrystal \cite{lifshitz_quasicrystals_2003,senechal_quasicrystals_2009,zilberberg_topology_2021}. Quasicrystals have been shown  to support fractal band structures \cite{bandres_topological_2016,zilberberg_topology_2021}. In our setting such non-commensurate orientations supports an infinite number of band gaps \cite{shah_tunneling_2021}. So it is at least conceivable that the band structure for a fixed irrational angle could also be a fractal. Here, we leave this as an interesting open question and rather focus on the orientation dependence of the spectrum.

\subsection{Resonant tunneling for arbitrary waveguide orientation}

From Fig.~\ref{fig:fig3}(a) it is clear that even though the number of band gaps is a discontinuous function of the orientation $\varphi$,  each given band gap can be viewed as defining a continuous function. In this section, we give  a theoretical underpinning to this observation by following the approach of Ref.~\cite{shah_tunneling_2021}.

\begin{figure*}
\centering
\includegraphics[width=2\columnwidth]{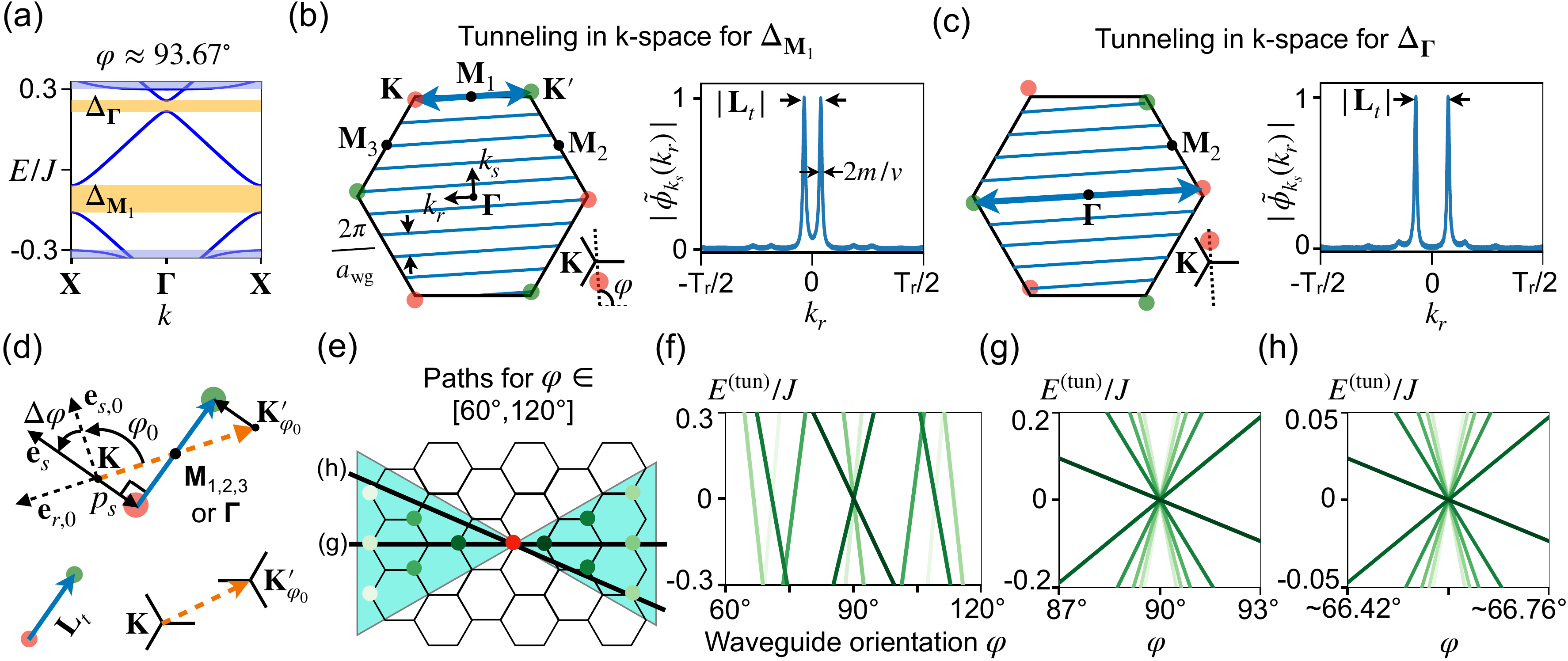}
    \caption{Waveguide band gaps induced by resonant tunneling. 
    (a) Waveguide band structure featuring the two most dominant waveguide band gaps. 
    (b, c) (left) Brillouin zone. Also sketched is the closed-loop support manifold (blue line) and the tunneling path  (between the blue arrows) for the wavefunction $\tilde{\boldsymbol{\psi}}_{\bar{k}_s}(\mathbf{k})$ for the lower-energy mode at the avoided crossing. The insets indicate that the positive velocity component is displaced from the $\mathbf{K}$ point according to Eq.~(\ref{eq:fourier_transform}). (right) Radial component of the wavefunction $\tilde{\phi}_{k_s}(k_r)$ as a function of the coordinate $k_r$ parameterizing the support manifold. 
    (d) Sketch of  a tunneling path giving rise to a band gap centered exactly at zero energy (orange), also indicated is the corresponding  waveguide orientation $\varphi_0$. After rotating the orientation  by an angle $\Delta\varphi$, the average quasimomenta (pink and green) and tunneling path (blue) are displaced.
    (e) Lattice in reciprocal space formed by the $\mathbf{K}$ and $\mathbf{K}'$ momenta. Each tunneling paths leading to a band gap centered at zero energy can be identified with an endpoint $\mathbf{K}'$ momentum once the other endpoint (red) has been fixed on the $\mathbf{K}$-sublattice.  The $\mathbf{K}'$ momenta leading to the first few shortest tunneling paths for orientations in the interval  $60^\circ\leq\varphi < 120^\circ$ (shaded region) are marked as green dots (darker markers indicate shorter paths). 
    (f-h) Resonant tunneling energy $E^{\rm (tun)}$ (band gap center) as a function of the waveguide orientation $\varphi$. Each line crossing zero energy and the corresponding zero-energy tunneling path is marked in (e). Sweeping the waveguide orientation $\varphi$ leads to a variation of the longitudinal quasi-momentum $p_s$, and hence of the tunneling energy $E^{\rm (tun)}$, see sketch in panel (d) and Eq.~(\ref{eq:tunn_ene_fixed_k_k'}). Zoom-ins about $\varphi_0 = 90^\circ$ and $\varphi_0 \approx 66.6^\circ$ are shown in (g,h). About all orientations $\varphi_0$ supporting zero-energy band gaps and  after an appropriate rescaling, the same pattern appears. The analytical results shown in panels (f-h) should be compared and partially explain the numerical results in Figs.~\ref{fig:fig3}(a-c).
    }
    \label{fig:fig4}
\end{figure*}

In the quasi-momentum representation, the wavefunction of the guided mode for an arbitrary (possibly irrational) $\alpha$ can be found using  the ansatz 
\begin{equation}
 \tilde{\boldsymbol{\psi}}_{\bar{k}_s}(\mathbf{k}) = \delta (k_s - \bar{k}_s) \, \tilde{\boldsymbol{\phi}}_{\bar{k}_s}(k_r). 
 \label{eq:edge_state_ansatz}
\end{equation}
Here, $k_s$ ($k_r$) is the quasimomentum in the direction longitudinal (transverse) to the domain wall, cf Fig.~\ref{fig:fig2}(a,b). For rational $\alpha$ (corresponding to translationally invariant waveguides, cf Eq.~(\ref{eq:alpha}))  this is just a Bloch-wave ansatz with waveguide quasimomentum $k=\bar{k}_s{\rm mod}(2\pi/a_{\rm wg})$.  More in general, $\bar{k}_s$ can be viewed as a label for the support manifold of the function $ \tilde{\boldsymbol{\psi}}_{\bar{k}_s}(\mathbf{k})$. This is a submanifold of the BZ parameterized by the radial quasi-momentum $k_r$. When the quasi-momentum $\mathbf{k}$ is taken inside the hexagonal Brillouin zone such submanifold transverse the BZ multiple times (for irrational $\alpha$ infinitely many times) defining a series of parallel lines, cf blue lines in the BZs depicted in Fig.~\ref{fig:fig4}(b,c). For rational $\alpha$, it is a closed manifold of length $T_r=A_{\rm BZ}a_{\rm wg}/2\pi$. We note that the Jackiw and Rebbi solution Eq.~(\ref{eq:fourier_transform}) represents a special example of the ansatz Eq.~(\ref{eq:edge_state_ansatz}) with the additional assumption that the wavefunction is  peaked close to the $\mathbf{K}$-point.  We mention in passing that a more general guided solution $ \tilde{\boldsymbol{\psi}}_{\bar{k}_s}(\mathbf{k})$ localized about a quasimomentum $\bar{\mathbf{k}}_{\bar k_s}$ that is not necessarily close to any high-symmetry point has been derived  in Ref.~\cite{shah_tunneling_2021}. Interestingly, this guided solution is unique (up to a reparametrization of $\bar{k}_s$)  and can be viewed as defining a single band comprising a right and a left propagating branch. As for the Jackiw and Rebbi solution, both the energy dispersion $E_{\bar{k}_s}$ and the classical quasi-momentum $\bar{\mathbf{k}}_{\bar k_s}$ depend only on  ${\rm sign}(m(x_r))$ but not on  $|m(x_r)|$. For $E_{\bar{k}_s}$ close to the middle of the bulk band gap, the classical quasimomentum $\bar{\mathbf{k}}_{\bar k_s}$ is close to a Dirac point recovering  the Jackiw and Rebbi solution. 

The emergence of the minigaps, forming the intricate pattern in Fig.~\ref{fig:fig3} is the consequence of the tunneling induced hybridization of  time-reversal-partner 
solutions localized about two opposite and possibly distant quasi-momenta, cf   Fig.~\ref{fig:fig4}(b,c). In section \ref{sec:mingaps} we have explained that for translationally invariant waveguides, the tunneling is resonant at the time-reversal-symmetric (waveguide) quasi-momenta $k=\Gamma$, and  $k=X$. For arbitrary waveguides, Ref.~\cite{shah_tunneling_2021} identified a more general condition: the two time-reversed solutions should have support on the same submanifold (but will be peaked at different $k_r$). This submanifold is itself time-reversal invariant. For irrational (rational) $\alpha$, the time-reversal invariant manifolds  pass  through one (two)  time-reversal invariant honeycomb lattice  quasi-momentum (quasimomenta),  $\mathbf{k}=\boldsymbol{M}_{i=1,2,3}$  or $\mathbf{k}=\boldsymbol{\Gamma}$, cf Fig.~\ref{fig:fig4}(d). In either scenario (rational or irrational $\alpha$) one can identify a dominant tunneling path, on which the overlap of the localized solutions is largest. This  path (indicated by the thick blue line in Fig.~\ref{fig:fig4}(b-d)) connects the two classical quasi-momenta  via a single time-reversal invariant quasimomentum. Two key intuitions allow to shed light on the intricate DOS pattern in Fig.~(\ref{fig:fig3}). First,  each continuous minigap is associated to a specific continuous (as a function of $\varphi$) tunneling path. Each such path (and the corresponding minigap) can be labeled with a time-reversal-invariant high-symmetry point (one of the $\boldsymbol{M}$-points  or the $\boldsymbol{\Gamma}$-point)   and the number of times it traverses each valley \cite{shah_tunneling_2021}, cf Fig.~\ref{fig:fig4}(a-c). Second, the different minigaps are of very different magnitudes. Quantifying these magnitudes is technically difficult as it requires to calculate the tail of the Jackiw and Rebbi solutions in quasi-momentum space,  going beyond the Dirac equation treatment \cite{shah_tunneling_2021}. However, for a  qualitative understanding of the DOS patterns observed in Fig.~\ref{fig:fig3}, it is enough to observe that there is a correlation between the size of a band gap and the length $|L_t|$ of the tunneling path, with longer tunneling paths leading to smaller band gaps, cf Fig.~\ref{fig:fig4}(a-c). Thus, only a finite number of the infinitely many  band gaps appears in a smeared out DOS, cf  Fig.~\ref{fig:fig3}. 

\subsection{Explanation of the fractal pattern}

Next, we focus on the guided modes spectrum  in the region centered about  $E=0$   where the interesting multiscale pattern  is observed, cf Fig.~(\ref{fig:fig3}). Our approach is centered  on special tunneling paths that are associated to  tunneling transitions that are resonant for exactly zero energy. In other words, they give rise to a minigap with center energy $E^{(\rm tun)}=0$ for an appropriate waveguide orientation $\varphi=\varphi_0$.
We refer to these tunneling paths as zero-energy tunneling paths. By analyzing their properties we will be able to show  that the same pattern for the center energies $E^{(\rm tun)}$ of the minigaps as a function of the angles repeats infinitely many times  about the axis $E=0$ for ever smaller energy and angle variations, overall, forming a self-similar fractal pattern.

About zero-energy we can determine the resonant tunneling energy $E^{(\rm tun)}$  using the Jackiw and Rebbi solution Eqs.~(\ref{eq:edge_state_energy},\ref{eq:fourier_transform}).  We preliminary note that for any arbitrary $\varphi$ the guided mode energy $E=0$ always has as  classical quasi-momentum $\bar{k}$  a Dirac point,   $\bar{k}=\mathbf{K}$ or $\bar{k}=\mathbf{K}'$. The two solutions at $\bar{k}=\mathbf{K}$ and $\bar{k}=\mathbf{K}'$ are time-reversal partners and the tunneling between them becomes resonant whenever a tunneling paths orthogonal to the waveguide orientation  connects these two high symmetry points. Geometrically, we can represent these zero-energy tunneling paths as described below. As a preliminary step, we draw the lattice that contains  all momenta corresponding to the $\mathbf{K}$ or the   $\mathbf{K}'$-points, cf Fig.~\ref{fig:fig4}(e). This is a honeycomb lattice in reciprocal space which has as lattice vectors the reciprocal lattice vectors $\mathbf{b}$.  Each zero-energy tunneling paths (once unfolded into the $\mathbb{R}^2$-plane) can be represented as a straight line connecting a fixed momentum  on the $\mathbf{K}$-sublattice  (e.g. the red point in Fig.~\ref{fig:fig4}(e)) with one of the infinite many  momenta on the $\mathbf{K}'$-lattice. In Fig.~\ref{fig:fig4}(e), we highlight the region of reciprocal space that hosts the zero-energy tunneling path for the waveguides orientations  $\varphi$ in the interval $60^\circ\leq\varphi<120^\circ$. In this region, we also highlight the  momenta on the $\mathbf{K}'$-lattice with different shades of green, encoding the length of the tunnelling path connecting these sites to the red dot on the $\mathbf{K}$-lattice. (As discussed above shorter paths correspond to larger band gaps.) Since the corresponding waveguide orientation  is orthogonal to the zero-energy tunneling path and, at the same time,  the reciprocal lattice is rotated by $90^\circ$ compared to the real space lattice, we conclude that  a waveguide supports a zero-energy band gap if and only if  moving in the longitudinal direction from a point on the $A$ sublattice one eventually crosses a point on the $B$ sublattice. While these particular set of orientations  are a dense subset of the  angles  corresponding to translationally invariant waveguides (and hence a dense subset of all orientations), in practice, only a finite and well spaced subset of orientations (those supporting the shortest zero-energy tunneling paths) will give rise to zero-energy band gaps  in a smeared out density of states, as shown in Fig.~\ref{fig:fig3}.

After having identified the waveguide orientations giving rise to large band gaps centered around zero energy, we  want to understand the structure of the density of states in the vicinity of these special orientations. We denote one such  orientation as $\varphi_0$. We also introduce the vectors $\mathbf{e}_{s,0}$ and $\mathbf{e}_{r,0}$ longitudinal and transverse to the domain wall [with the same conventions as in Fig.~\ref{fig:fig1}(c)], and the momentum $\mathbf{K}'_{\varphi_0}$ on the $\mathbf{K}'$-lattice such that the vector $\mathbf{K}'_{\varphi_0}-\mathbf{K}$ is orthogonal to $\mathbf{e}_{s,0}$. Simple geometrical considerations, cf Fig.~\ref{fig:fig4}(d), show that the linear dispersion of the Jackiew and Rebbi solution Eq.~(\ref{eq:edge_state_energy}) gives rise to a linear dependence  of the resonant tunneling energy $E^{\rm (tun)}$ on the waveguide orientation $\Delta\varphi=\varphi-\varphi_0$ counted off from $\varphi_0$,
\begin{equation}
    E^{\rm (tun)}(\Delta\varphi) = vp_s (\Delta\varphi) = \frac{v L_0}{2} \sin (\Delta\varphi) 
     \approx  \frac{v L_0}{2}\Delta\varphi.
    \label{eq:tunn_ene_fixed_k_k'}
\end{equation}
Here,   $L_0$ is the projection of the  zero-energy tunneling path along the $r$-axis, $L_0=(\mathbf{K'}_{\varphi_0}-\mathbf{K})\cdot \mathbf{e}_{r,0}$ (equivalently, this is the length of the zero-energy tunneling path multiplied by the appropriate sign). 
This formula allows to predict the slope of the band gap center $E^{\rm (tun)}(\varphi)$  for each band gap that crosses zero energy in terms of the  zero-energy tunneling path or, equivalently, the corresponding $\mathbf{K}'$ and $\mathbf{K}$ end momenta. As explained above the width of the band gap is proportional to the tunneling length and, in practice, only the shortest tunneling paths will give rise to appreciable band gaps. The band gap centers for the shortest zero-energy tunneling paths (those corresponding to the $\mathbf{K}'$ momenta highlighted as green dots in Fig.~\ref{fig:fig4}(e)) calculated using Eq.~(\ref{eq:tunn_ene_fixed_k_k'})  are shown in Fig.~\ref{fig:fig4}(f). We note that several band gaps merge at the same origin on the zero-energy axis as previously observed for the density of states in Fig.~(\ref{fig:fig3}). This simply reflects that  there are (infinitely) many $\mathbf{K}'$ momenta  on every line that also  crosses  a $\mathbf{K}$ momentum, cf Fig.~\ref{fig:fig4}(e). 
Geometrical considerations shows that  if we label these $\mathbf{K}'$ momenta with an integer label $n\in\mathbb{Z}$ choosing $n=0$ for the shortest zero-energy tunneling path (for a fixed $\varphi_0$) the other labels can be chosen such that 
\begin{equation}
    L_{\varphi_0}^{(n)}\equiv(\mathbf{K'}^{(n)}_{\varphi_0}-\mathbf{K})\cdot \mathbf{e}_{r,0} = (3n+1) L_{\varphi_0}^{(0)}
    \label{eq:triangular_lattice_property},
\end{equation}
with $n\in \mathbb{Z}$ see Appendix~\ref{app:fractal_property}. In other words the  ratio $L_{\varphi_0}^{(n)}/L_{\varphi_0}^{(0)}=3n+1$ is   independent on $\varphi_0$.  This relation can be rephrased more generally as a constraint on the coordinates of the sites of a honeycomb lattice, see Appendix~\ref{app:fractal_property}. In our setting, the parameters  $L_{\varphi_0}^{(n)}$ set the slope of the tunneling energies $E_{\rm tun}$,  given by Eq.~(\ref{eq:tunn_ene_fixed_k_k'})  with $L_0=L_{\varphi_0}^{(n)}$. Thus, the local  pattern of the resonant tunneling energies $E_{\rm tun}$ about a zero-energy tunneling orientation $\varphi_0$ is always the same, with the angle $\varphi_0$ only fixing the scale via  $L_{\varphi_0}^{(0)}$, cf  Fig.~\ref{fig:fig4}(g-h).  Overall, in the ideal limit of an infinite system without dissipation, this induce a self-similar fractal pattern for the tunneling energies as a function of the waveguide orientation $\varphi$.

Until now we have analysed the fractal pattern for the  tunneling energies as a function of the waveguide orientation. Next, we discuss how a similar pattern is imprinted in the DOS. We note that our analysis so far  only partially explains the empirical observation that exactly the same pattern is  observed in  the DOS at two different scales, cf Figs.~\ref{fig:fig3}(b,c). In particular, it  accounts for the slopes of the band gaps but not for their widths. While we  expect that the widths (set by the underlying tunneling rate) will be  proportional to the length of the zero-energy tunneling path, the dependence will in general be non-linear. For this reason, zoom-ins of the DOS about different $\varphi_0$ may look substantially different even after accounting for the different scale factors. Nevertheless,
they will all display a series of band gaps of varying width merging at zero energy. Thus, the overall global dependence of the DOS  on the waveguide orientation inherits the self-repeating fractal nature of the resonant tunneling energies.

\begin{figure*}
\centering
\includegraphics[width=2\columnwidth]{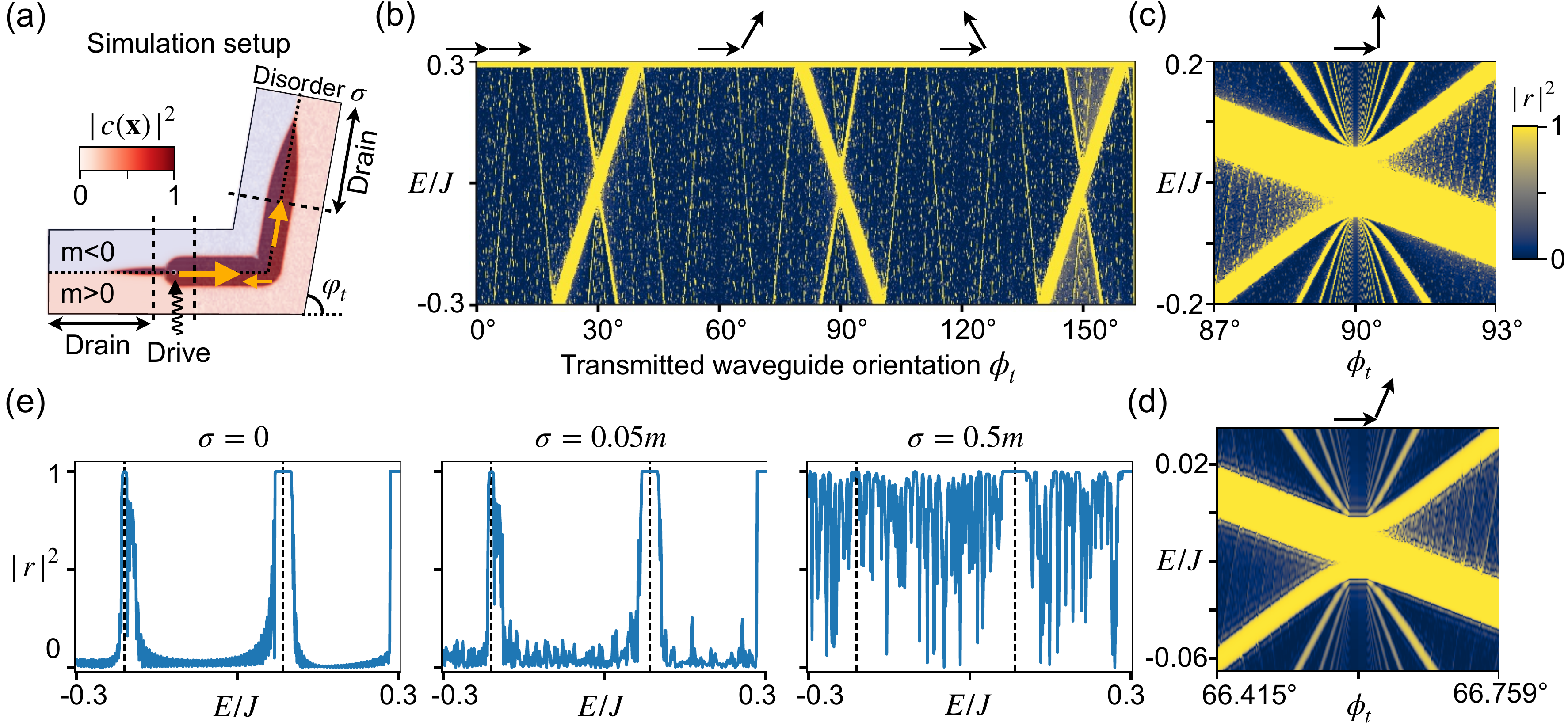}
\caption{Fractal signatures in the backscattering at a corner. 
(a) Simulation setup to determine the backscattering rate $\vert r \vert ^2$. The steady state distribution (see color bar) has decaying tails at the incident and transmission drains. 
(b) Backscattering rate as a function of energy and the final waveguide orientation $\varphi_t \in [0^\circ, 165^\circ]$ for the setup in (a). 
(c, d) Magnified image of (b) near $\varphi=90^\circ$ and $\varphi=66.6^\circ$ featuring the fractal pattern in analogy to the Figs.~\ref{fig:fig3}(b,c). 
(e) Reflection spectra for $\varphi_t \approx 97.05^\circ $ ($p=-3,q=14$) for increasing values of the disorder $\sigma$. The dashed vertical lines indicate the central energy of the edge-band gaps.
}
\label{fig:fig5}
\end{figure*}

\section{Fractal transmission  at a sharp corner}
\label{sec:fractal_backscattering}

Next, we analyze a transport scenario that would allow to directly observe the fractal features analyzed so far.
We consider a setup formed by connecting two waveguides of different orientations via a sharp corner, see Figs.~\ref{fig:fig1} and \ref{fig:fig5}(a). We are mainly interested in the  dependence of the transmission on the waveguides orientations. We mention in passing that in the hard domain-wall scenario considered here, the exact location of  the domain wall relative to the lattice can also influence the transmission quantitatively (see Appendix~\ref{app:deviation_smooth_envelope}). However, no qualitative change is observed indicating that the smooth envelope approximation is sufficient to capture the essential physics of the problem.

The simulation setup is shown in Fig.~\ref{fig:fig5}(a). We drive two  sites on the initial waveguide with the same amplitude and frequency (energy), but a suitable phase difference, to unidirectionally excite the edge channel towards the corner (see Appendix~\ref{app:sim_setup}). At the corner, the wave is partially backscattered. Eventually,  the transmitted and the reflected waves are  absorbed in two different  drains. To prevent any undesirable backscattering at the drain-waveguide interface, the dissipation rate is increased very slowly in the drains, see Appendix~\ref{app:sim_setup} for more details. Therefore, we need large drains, which results in a tight-binding model of $5 \times 10^5$ lattice sites. We use sparse matrices to store and perform linear algebra operations on such large dimensional matrices. 
We derive the steady state backscattering rate $\vert r\vert ^2$ from the ratio of the absorbed intensities in the transmission and absorption drains, see Appendix~\ref{app:sim_setup}. We choose the first waveguide to support a zig-zag domain wall ($\varphi = 0^\circ$) because this configuration does not feature a band gap for the parameters considered here. In this way, an excitations of frequency inside the bulk band gap  can always be injected in the first waveguide  and will only  be backscattered if its frequency is in the band gap of the second waveguide.  A more complex scenario with a different initial waveguide orientation is described in the Appendix~\ref{app:armchair_to_all_angles}. The backscattering rate $\vert r\vert ^2$ as a function of the energy $E$ and the final waveguide orientation $\varphi_t$ is shown Fig.~\ref{fig:fig5}(b). As expected, the backscattering rate spikes up whenever the final waveguide band structure has a band gap at that energy. Moreover, Fig.~\ref{fig:fig5}(c,d) features a similar fractal pattern as that in Fig.~\ref{fig:fig3}. To mimic a realistic experimental systems, we introduce disorder in the onsite potentials of the lattice sites. We observe that the fractal pattern is roughly intact up to disorder levels an order of magnitude smaller than the bulk band gap $2m$, see Fig.~\ref{fig:fig5}(e).

\section{Discussion and outlook}

Our investigation of the orientation dependence of the density of state and transmission in valley Hall waveguides has focused on a specific case study, a simple extension of the graphene tight-binding model. Nevertheless, the physics described here is robust and extends beyond the validity of this specific model. Importantly, we have relied on it only for our numerical calculations. On the other hand, we have explained the repeating pattern featuring multiple waveguide's band gaps merging in the middle of the bulk band gap  solely based  on  the effective Dirac Hamiltonian  Eq.~(\ref{eq:dirac_hamiltonian}) and the position of the Dirac cones in the crystal BZ (which is fixed by the underlying space group). We can conclude that a qualitatively similar pattern will be observed in any valley Hall waveguide with  the suitable space group symmetries of the bulk crystal structure (at least $p6$ or $p31m$ before introducing the gap-opening perturbation, see e.g. \cite{shah_topologically_2022}). More precisely, the centers of the band gaps are captured by Eq.~(\ref{eq:dirac_hamiltonian}) while the widths depend on the tails of the guided modes and, thus, on the microscopic wave equations.  Reconfigurable implementations of topological waveguides \cite{fleury_floquet_2016,cheng_robust_2016,zhang_topological_2018,zhao_non-hermitian_2019,darabi_reconfigurable_2020,tian_dispersion_2020} are ideally suited to verify our predictions.

More generally, we expect that the physics investigated, here,
to be relevant for other topological waveguides obtained using a variety of schemes  based on the geometric engineering of crystal structures supporting   Dirac cones  \cite{shah_topologically_2022}. In this more general setting, the counter-propagating guided modes are usually described  within a smooth-envelope approximation that does not capture their coupling via their tails. This coupling becomes significant in the regime of tight transverse confinement/large band gaps, $m/va\gtrsim 1$. In this regime, we  expect that  tunneling induced band gaps can be observed. As for the valley Hall waveguides investigated in this work,  the number of band gaps will depend discontinuously on the waveguide orientation leading to qualitatively different but similarly intricated patterns with merging band gaps for the band structure/transmission vs waveguide orientation. 

We are confident that our investigation provides a significant contribution toward the general goal of achieving efficient routing of photons or phonons along arbitrary paths in densely-packed integrated circuits.

\vspace{2mm}

\noindent\textbf{Acknowledgements}\\ 
T.S. acknowledges support from the European Union’s Horizon 2020 research and innovation programme under the Marie Sklodowska-Curie grant agreement No. 722923 (OMT).  F.M. acknowledges support from the European Union’s Horizon 2020 Research and Innovation program under Grant No. 732894, Future and Emerging Technologies (FET)-Proactive Hybrid Optomechanical Technologies (HOT).


\bibliography{references}
\clearpage
\onecolumngrid

\appendix

\section{Calculation of density of states}
\label{app:density_of_states}
\begin{figure}
\centering
\includegraphics[width=1\columnwidth]{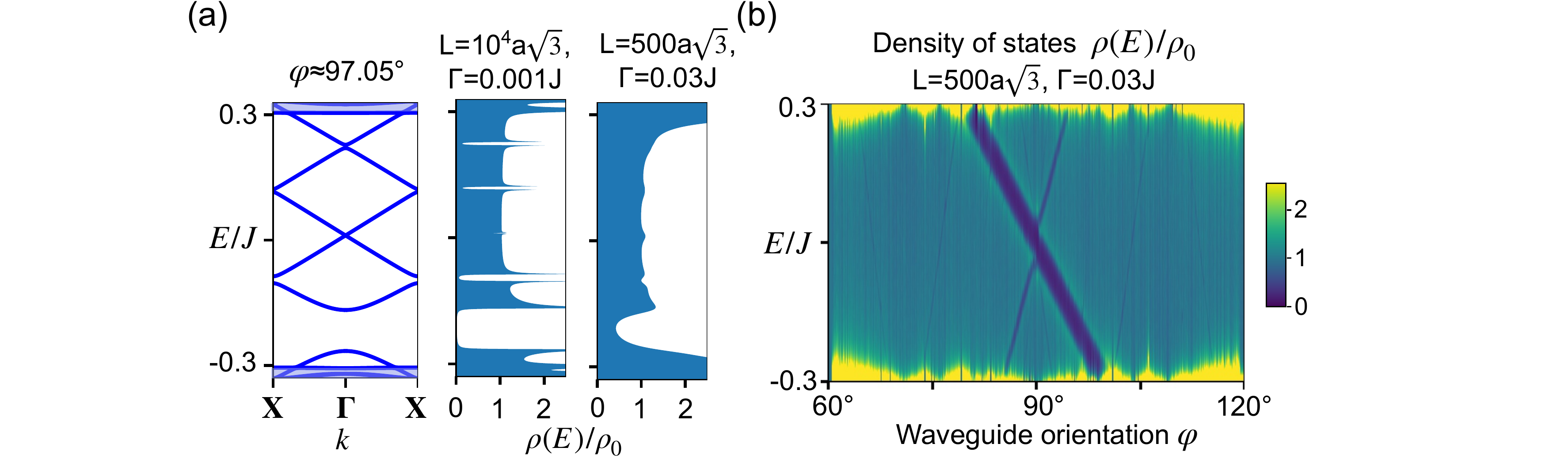}
\caption{(a) Numerically evaluated waveguide band structure with the orientation $\varphi \approx 97.05 ^\circ$ ($p=-3,q=14$), and the corresponding edge mode density of states for two different sets of system size $L$ and dissipation rate $\Gamma$. (right) For a smaller system size and higher dissipation rate, the edge-band gaps with smaller width are unresolved.
(b) Density of state pattern with smaller system size and larger dissipation rate. Only the two most dominant edge-band gaps can be resolved (compare with Fig.~\ref{fig:fig3} of the main text).
}
\label{fig:app_dos}

\end{figure}
In this section, we describe how to evaluate the density of states of the edge mode, which is presented in the Fig.~\ref{fig:fig3} of the main text. In addition, we point out that the fractal structure of the density of states is more discernible for large system sizes and low dissipation.

Consider a topological waveguide of longitudinal length $L$ and orientation $\varphi$. It can fit $N_{\rm wg}\sim L/\vert a_{\rm wg}(\varphi) \vert$ waveguide unit cells. Hence, the waveguide edge modes can be represented with a set of discrete quasi momenta $k_s$ with separation $\Delta k_s = 2\pi /L$. For a system with dissipation rate $\Gamma$, the density of states is defined as
\begin{equation}
    \rho (E) = \frac{2}{2\pi} \sum _{n,k_s} \frac{\Gamma}{(E - E_{n,k_s})^2 + (\Gamma /2)^2},
    \label{eq:dos}
\end{equation}
where $E_{n,k_s}$ is the energy of the $n^{\rm th}$ folded band ($n=\{0,1,...,N-1 \}$) in the waveguide band structure, and the numerator $2$ represents the two counter-propagating edge modes in the bulk band gap. Mathematically, Eq.~(\ref{eq:dos}) represents the sum of multiple Lorentzian distributions of width $\Gamma$ and separation $\Delta E \approx v_{n,k_s} \Delta k_s$ ($v_{n,k_s}=dE_{n,k_s}/dk_s$). Hence, only those edge-band gaps can be resolved via $\rho (E)$, whose width $\Delta$ is larger than the dissipation ($\Delta > \Gamma$) and the separation between Lorentzians ($\Delta > \Delta E$), cf Fig.~\ref{fig:app_dos}. 

For a large system size $v \Delta k_s \ll \Gamma$, we can approximate the summation over $k_s$ in Eq.~(\ref{eq:dos}) with an integral to obtain
\begin{equation}
    \rho (E) = \frac{2}{2\pi \Delta k_s} \sum _{n} \int  dE_{n,k_s} \frac{\Gamma /v_{n,k_s}}{(E - E_{n,k_s})^2 + (\Gamma /2)^2}.
    \label{eq:dos_integral_form}
\end{equation}
Therefore, for the ideal scenario of no edge-band gaps and a linear band dispersion $v_{n,k_s}=v$, the density of states is given by $\rho_0 = L/(\pi v)$. Note that the group velocity $v_{n,k_s}$ is close to zero at the boundaries of the edge-band gap, hence $\rho (E)$ spikes up at these energy values with decaying tails inside the gap.

\section{Proof of the fractal property of the triangular lattice}
\label{app:fractal_property}
\begin{figure}
\centering
\includegraphics[width=1\columnwidth]{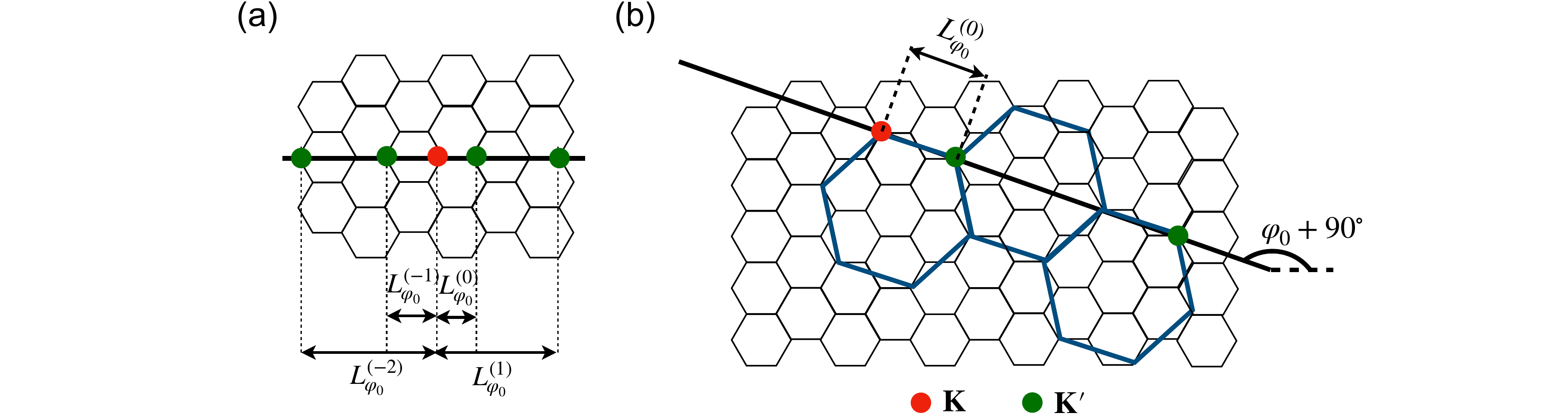}
\caption{Geometrical proof of the fractal property of the triangular lattice. 
(a) For a horizontal line (black) ($\varphi_0=90^\circ$) passing through the $\mathbf{K}$ (red) and $\mathbf{K'}$ (green) points, the zero-energy tunneling paths are given by $L_{\varphi_0}^{(n)} = (3n+1) L_{\varphi_0}^{(0)}$. 
(b) For an arbitrary line with orientation $\varphi_0$, the proof is identical to the case of $\varphi_0=90^\circ$. This is because the line passes through the corners of the larger unit cell (blue hexagons) of the reciprocal space similar as in (a). 
}
\label{fig:app_fractal}

\end{figure}
In this section, we prove the geometrical property of the triangular lattice, that led to the fractal structure of the tunneling energy as a function of the waveguide orientation, cf Eq.~(\ref{eq:triangular_lattice_property}) of the main text. More specifically, we prove the following property: 

Consider a straight line on the lattice in reciprocal space (perpendicular to the waveguide orientation $\varphi_0$) passing through the two points $\mathbf{K}$ and $\mathbf{K'}$. Then, the ratio of the lengths of all the zero-energy tunneling paths $L_{\varphi_0}$ directed from a fixed $\mathbf{K}$ point to an arbitrary $\mathbf{K'}$ point along this line is independent of the waveguide orientation $\varphi_0$, and is given by $L_{\varphi_0}^{(n)} = (3n+1) L_{\varphi_0}^{(0)}$. Here, $n$ is an integer and $L_{\varphi_0}^{(0)}$ is the shortest length from $\mathbf{K}$ to $\mathbf{K'}$.

It is intuitive to look at the proof geometrically, see Fig.~\ref{fig:app_fractal}. For a horizontal line $\varphi_0 = 90^\circ$, the above property is obvious, cf Fig.~\ref{fig:app_fractal}(a). For an arbitrary $\varphi_0$, consider a hexagon with the \textit{shortest} zero-energy tunneling path $L_{\varphi_0}^{(0)}$ as its side, see Fig.~\ref{fig:app_fractal}(b). Due to the $60^\circ$ rotational symmetry of the lattice, this hexagon must be a (larger) unit cell of the lattice in the reciprocal space. By observing the lattice from the point of view of the larger unit cell, the proof is identical to the simpler case for $\varphi_0 = 90^\circ$. Note that there cannot be a pair of $\mathbf{K}$ and $\mathbf{K'}$ points on the line inside the larger unit cell. If that would be the case, it would contradict our premise that the hexagon is created with the shortest zero-energy tunneling path $L_{\varphi_0}^{(0)}$ as its side.

\section{Deviations from smooth-envelope approximation}
\label{app:deviation_smooth_envelope}
\begin{figure}
\centering
\includegraphics[width=1\columnwidth]{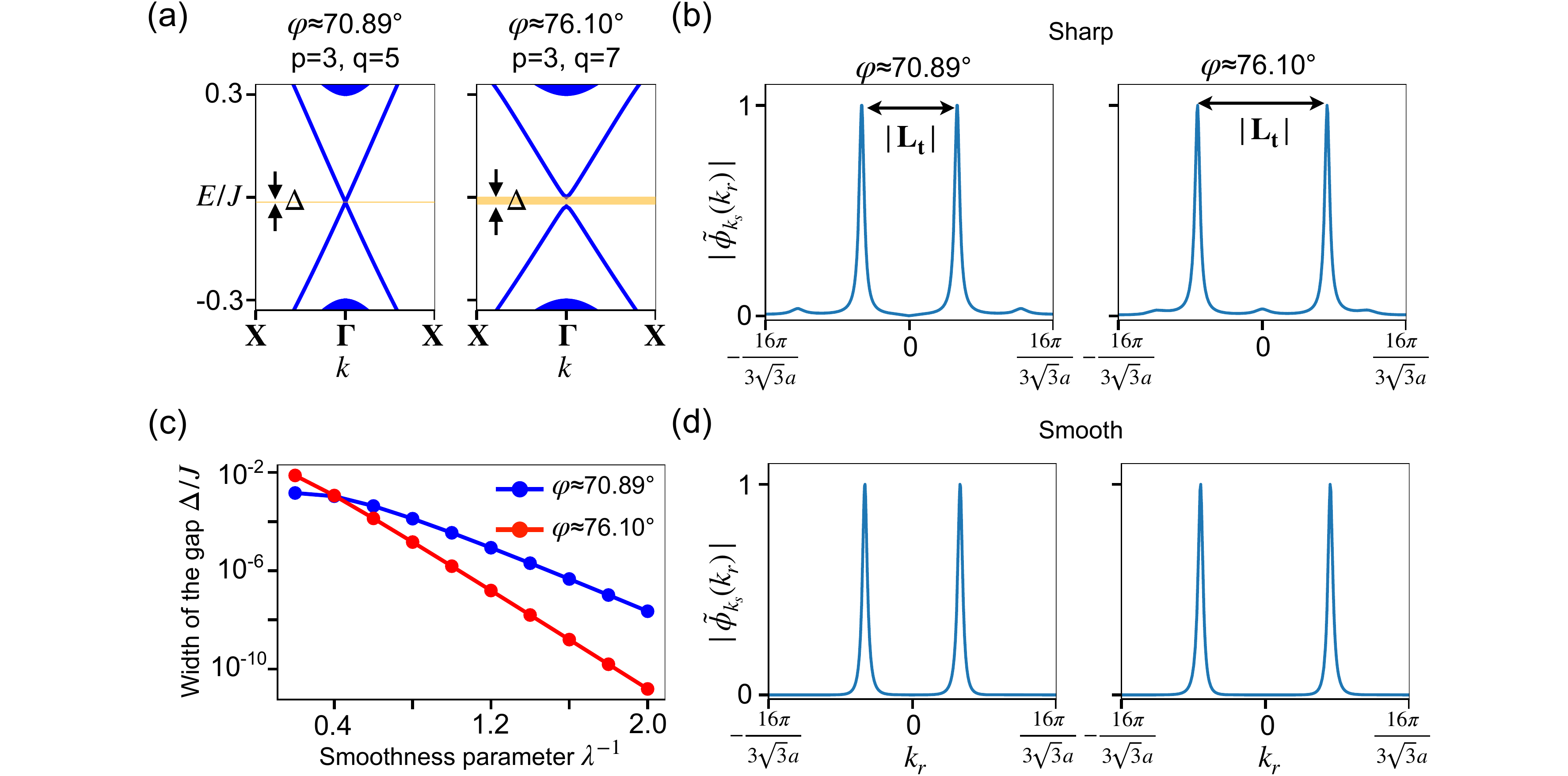}
\caption{A shorter tunneling path corresponds to a larger edge-band gap in the \textit{smooth-envelope regime}. 
(a) Waveguide band structures for two different orientations. 
(b) Fourier transform for the lower energy mode at the avoided crossing for the two cases in (a). For a sharp transition between the two domains, the case with the shorter tunneling path has a smaller edge-band gap (contrary to the expectation). 
(c) Edge-band gap width $\Delta$ as a function of the smoothness parameter $\lambda^{-1}$ of the domain transition. The mass parameter $m$ varies across the domain wall according to the relation $m(x_r) = -m_{bk} \tanh (\lambda x_r/a)$. For a smoother transition, the case with the shorter tunneling path has a larger edge-band gap (as expected). 
(d) Same as (b) for the smooth domain transition $\lambda^{-1}=2$.
}
\label{fig:app_smooth_envelope_1}

\end{figure}

\begin{figure}
\centering
\includegraphics[width=1\columnwidth]{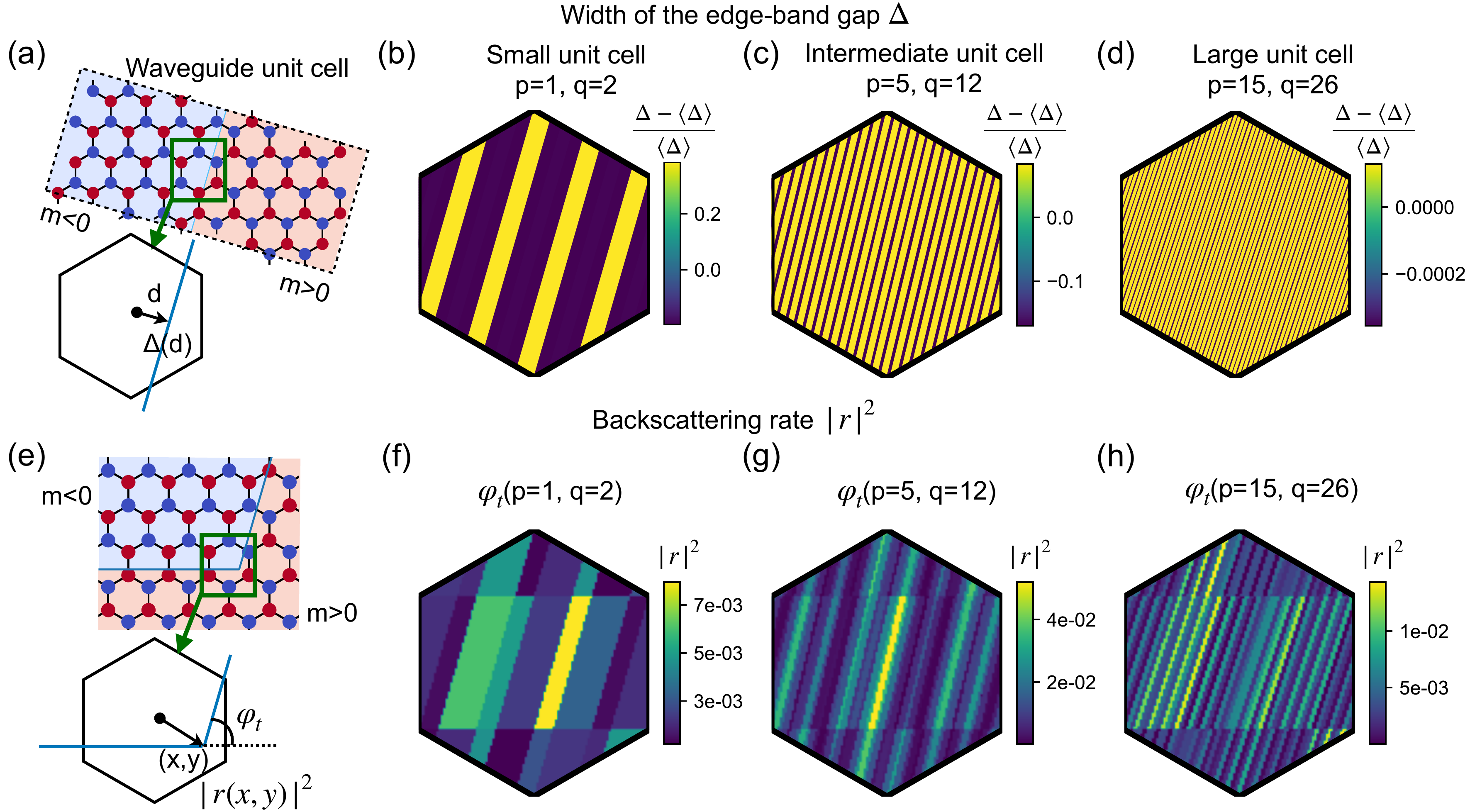}
\caption{Dependence of the (a-d) edge-band gap width $\Delta$ and (e-h) the backscattering rate $\vert r \vert ^2$ on the position of the domain wall and the corner, respectively. (a) and (e) The Hamiltonian depends on the configuration of the sublattices A and B in the immediate vicinity of the domain wall. (b-d) and (f-h) The quantities $\Delta$ and $\vert r \vert ^2$ are constant within the different stripes in the hexagon bulk unit cell. The number of stripes increases with the waveguide lattice constant $a_{\rm wg}$.
}
\label{fig:app_smooth_envelope_2}
\end{figure}

The smooth-envelope approximation assumes that the excitation varies over a much longer length scale than the lattice spacing. Hence, any sudden modification of the geometry, e.g. a sharp domain wall interface in the waveguide unit cell or a sharp turn between two dissimilar waveguides, is not taken into account by this approximation.
Below, we highlight two such scenarios where the observations deviate from those predicted within the smooth-envelope approximation.

According to the smooth-envelope approximation, the edge-band gaps corresponding to a shorter tunneling path should have a larger width. However, a detailed comparison between the Figs.~\ref{fig:fig4}(f) and \ref{fig:fig3}(a) features deviations from this expectation. In Fig.~\ref{fig:app_smooth_envelope_1}, we show that the deviation arises indeed due to a sharp interface between the two distinct domains. We consider the two different edge-band gaps corresponding to different tunneling paths, and show that the edge-band gap with longer tunneling path has a smaller width for increased smoothness of the domain wall interface, see Fig.~\ref{fig:app_smooth_envelope_1}(c). In addition, the Fourier transforms for the case of a sharper interface features kinks, which disappears for a smoother interface, cf Fig.~\ref{fig:app_smooth_envelope_1}(b,d).

An alternative consequence of the sharp domain wall interface (sharp turn) is that the width of the edge-band gap (backscattering rate) depends on the exact position of the domain wall interface (corner) with respect to the underlying lattice, see Fig.~\ref{fig:app_smooth_envelope_2}. For the case of a sharp domain wall interface in the waveguide unit cell, the Hamiltonian is dependent on the configuration of the A and B sublattices at the immediate vicinity of the domain wall interface, cf Fig.~\ref{fig:app_smooth_envelope_2}(a). 
Therefore, one can investigate the width of the edge-band gap as a function of the distance $d$ of the interface from the center of the hexagon unit cell, see Figs.~\ref{fig:app_smooth_envelope_2}(b-d). We observe that the hexagon can be divided into many constant-width stripes oriented parallel to the waveguide. The stripes correspond to the same waveguide Hamiltonian, and the number of stripes increases with the waveguide lattice constant $a_{\rm wg}$. 
Similar stripes can also be observed for the backscattering rate as a function of the corner position within the hexagon, see Figs.~\ref{fig:app_smooth_envelope_2}(e-h). However, for this case, the location of the stripes depend on the orientation of both the incident and transmission waveguides.

\section{Simulation setup to evaluate the backscattering rate}
\label{app:sim_setup}
\begin{figure}
\centering
\includegraphics[width=1\columnwidth]{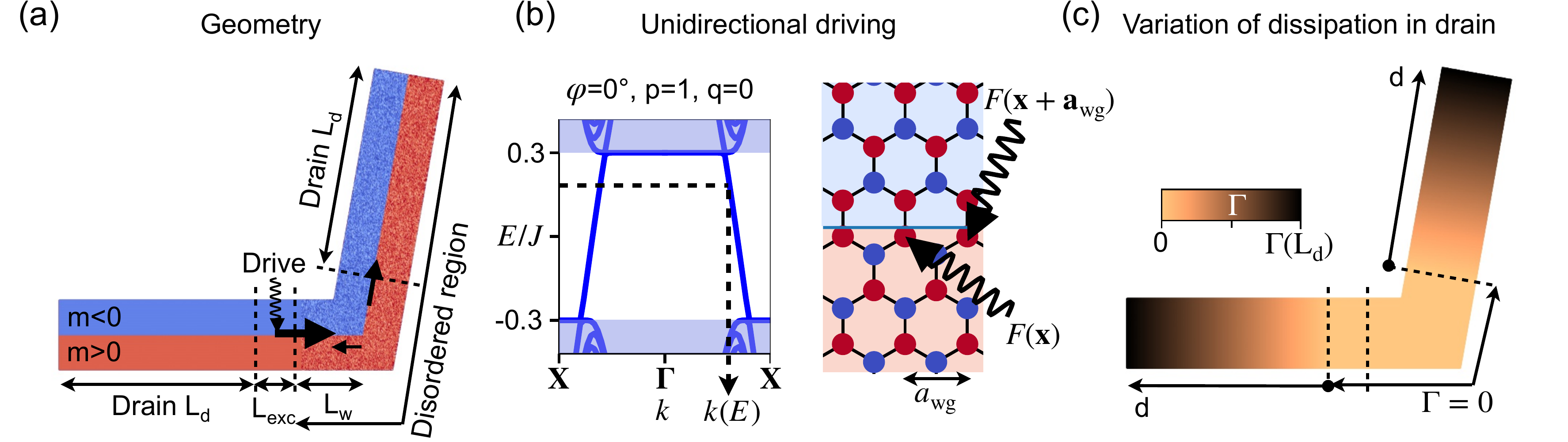}
\caption{Tight-binding simulation setup to evaluate the backscattering rate $\vert r \vert ^2$. 
(a) Geometry of the setup featuring two distinct domains (red and blue), two drains (length $L_{\rm d}$), a translationally invariant region around the driven sites (length $L_{\rm exc}$), and the disordered region containing the two waveguides of different orientations (each of length $L_{\rm w}$) and the transmitted drain. 
(b) (left) Waveguide band structure for the zigzag orientation. For a fixed drive energy, we obtain the quasimomentum $k(E)$ (dashed line) of the mode that is traveling away from the corner. (right) The unidirectional driving of the incident wave towards the corner is achieved by driving two translationally invariant sites with the phase difference that depends on $k(E)$. 
(c) Variation of the dissipation rate $\Gamma$ in the drain regions. 
Parameter values considered in the simulation: $L_{\rm d}=1000a\sqrt{3}, \, L_{\rm exc}=200a\sqrt{3}, \, L_{\rm w}=300a\sqrt{3}, \, \Gamma_{\rm max} = 10\vert J\vert, \, \lambda_d = 5 \times 10^{-4}$.
}
\label{fig:app_simulation_setup}

\end{figure}

In this section, we present the details about the simulation setup, that is used to evaluate the backscattering rate $\vert r \vert ^2$ in Fig.~\ref{fig:fig5} of the main text. 

The Hamiltonian $\hat{H}$ of the setup, see Fig.~\ref{fig:app_simulation_setup}(a), is given by Eq.~(\ref{eq:real_space_hamiltonian}) of the main text with opposite signs of the mass term $m$ in the two domains. A setup containing $N$ lattice sites is represented by a $N \times N$ dimensional Hamiltonian matrix $\hat{H}$. For the general case of a site-dependent drive $F_n e^{-iEt}$ ($n=\{0,1,..., N-1\}$ is the site index) at frequency (energy) $E$, and a site-dependent dissipation rate $\hat{\Gamma}$ ($N \times N$ diagonal matrix), the steady state solution $c_n$ at the $\rm n^{th}$ site can be written as
\begin{equation}
    c_n = \sum_{m} \chi_{nm} F_m e^{-iEt}, 
    \, \, \text{where }
    \hat{\chi}[E] = \left[ i(\hat{H} - E \, \hat{\mathbb{I}}) + \frac{\hat{\Gamma}}{2} \right]^{-1}.
    \label{eq:steady_state}
\end{equation}
Here, $\hat{\chi}[E]$ is the susceptibility matrix, and $\hat{\mathbb{I}}$ is the identity matrix. Below, we describe the details of the drive $F_n$, the dissipation rate $\Gamma_n$, and the relation between the steady state solution $c_n$ and the backscattering rate $\vert r \vert ^2$.

We want the harmonic drive to excite the edge channel unidirectionally towards the corner. At the fixed drive energy $E$, there are two possible counter-propagating edge modes in the band structure of the initial waveguide, cf Fig.~\ref{fig:app_simulation_setup}(b). The excited wave can be made to travel \textit{only} towards the corner if the overlap of the drive with the waveguide mode $\psi_n$ traveling away from the corner is zero i.e. 
\begin{equation}
    \sum _n F_n^* \psi_n = 0.
    \label{eq:unidirectional_excitation}
\end{equation}
At the two translationally invariant sites (indicated by the indices $m=\{0,1 \}$) in the initial waveguide at positions $\mathbf{x}+m\,\mathbf{a}_{\rm wg}$, the waveguide mode that is travelling away from the corner is proportional to $\psi_m \propto e^{imk(E) a_{\rm wg}}$. Here, $k(E)$ is the invariant quasimomentum at the energy $E$, see Fig.~\ref{fig:app_simulation_setup}(b). Thus, Eq.~(\ref{eq:unidirectional_excitation}) is satisfied if we drive two sites with the phase $F_m = (-1)^m \psi_m$. Note that the above reasoning behind the unidirectional excitation is relied on the assumption of the translational invariance of the initial waveguide. Thus, the driven sites are located sufficiently far from the beginning of the initial drain and the disordered onsite potential region, cf Fig.~\ref{fig:app_simulation_setup}(a).

The dissipation rate $\Gamma_n$ is non-zero only at the drains. Within each drain, the dissipation rate is given by $\Gamma (d) = \Gamma_{\rm max} \tanh \left(\lambda_d d/a \right)$, where $d$ is the distance of the site from the waveguide-drain interface, cf Fig.~\ref{fig:app_simulation_setup}(c). The smoothness parameter $\lambda$ is chosen to be sufficiently small to prevent undesirable backscattering at the drain-waveguide interface.

The backscattering rate $\vert r \vert ^2$ is given by the ratio of the absorbed intensity in the transmission drain to the total absorbed intensity in both the drains i.e. 
\begin{equation}
    \vert r \vert ^2 = \frac{\sum_{n \in \text{trans. drain}} \Gamma_n \vert c_n \vert ^2}{\sum_{n \in \text{both drains}} \Gamma_n \vert c_n \vert ^2}.
    \label{eq:backscattering_rate}
\end{equation}

\section{Backscattering rate with initial waveguide at armchair orientation}
\label{app:armchair_to_all_angles}

\begin{figure}
\centering
\includegraphics[width=1\columnwidth]{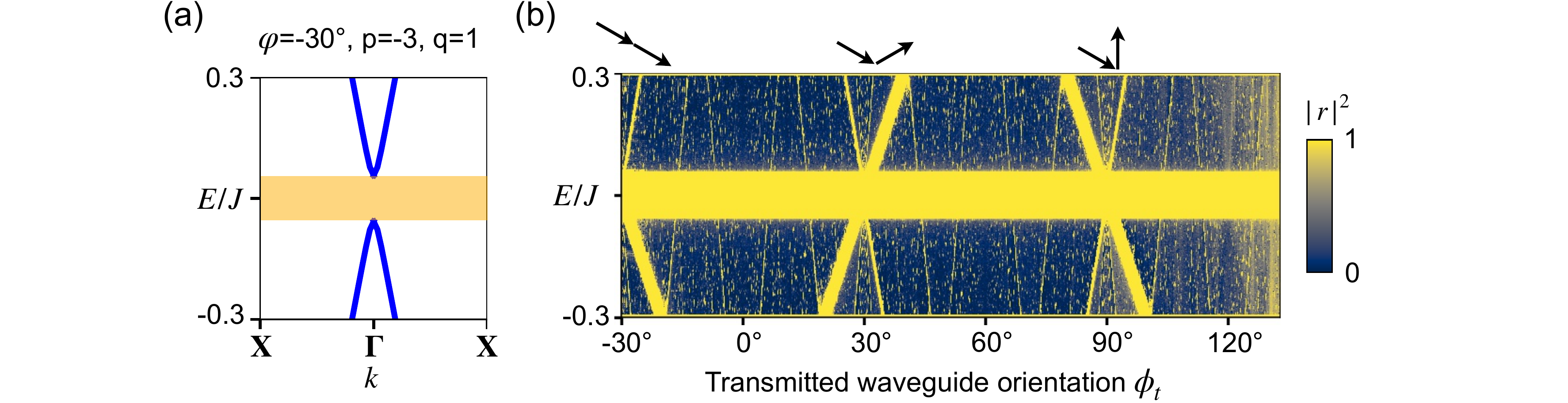}
\caption{Backscattering rate $\vert r \vert ^2$ with initial waveguide at the armchair orientation $\varphi _i = -30^\circ$. 
(a) Waveguide band structure for the armchair orientation featuring an edge-band gap near zero energy. 
(b) Backscattering rate $\vert r \vert ^2$ as a function of the energy $E$ and the final waveguide orientation $\varphi _f$ for the fixed initial waveguide at the armchair orientation.
}
\label{fig:app_armchair_to_all_angles}

\end{figure}

In this section, we present the backscattering rate $\vert r \vert ^2$ as a function of the energy $E$ and the transmission waveguide orientation $\varphi_t$ for the fixed initial waveguide at the armchair orientation $\varphi_i = -30^\circ$, see Fig.~\ref{fig:app_armchair_to_all_angles}(b). We notice that the pattern is identical to Fig.~\ref{fig:fig5} of the main text, except that there is an additional edge-band gap near zero energy due to the fixed initial waveguide, cf Fig.~\ref{fig:app_armchair_to_all_angles}(a).

\end{document}